# RCL: A Retrieval-Confidence Layer for Detecting Insufficient Context in Enterprise Retrieval-Augmented Code Generation

Chandra Mohan Ravuri

*Independent Researcher*

**Abstract**

Retrieval-Augmented Generation (RAG) for code generation has been extensively studied on public repositories, where a large language model's parametric knowledge, acquired during pretraining, often compensates for imperfect or incomplete retrieval. This compensating effect breaks down in enterprise codebases, where private APIs, internal frameworks, and undocumented team conventions fall entirely outside any model's pretraining distribution. Recent work on private-library code generation shows that even oracle (perfect) retrieval does not eliminate errors, but locates the remaining failures downstream, in API usage; separately, confidence-gated retrieval has been studied for general open-domain question answering using model-internal confidence. Neither addresses whether retrieval itself was structurally sufficient for a private-code query before generation begins. We introduce RCL (Retrieval-Confidence Layer), a lightweight module inserted between retrieval and generation that combines a call-graph-derived structural coverage score with a novelty score estimating a query's dependence on knowledge outside the model's prior, to detect insufficient retrieval before generation occurs. When the combined confidence score falls below a calibrated threshold, RCL triggers a targeted follow-up retrieval pass or labels the eventual output for human review, rather than allowing the system to generate silently against incomplete context. We describe RCL's architecture, formalize its scoring functions, and propose an evaluation methodology using a private-code benchmark constructed by injecting synthetic internal APIs and naming conventions into open-source Java repositories, simulating the enterprise condition without requiring access to proprietary code. We report results (Section 7) comparing RCL against similarity-only retrieval baselines on generation correctness. Our position is that retrieval sufficiency, assessed structurally rather than from model-internal confidence, is a distinct and currently underaddressed signal for building safer code-generation systems in private, enterprise settings.

## 1. Introduction

Large language models (LLMs) integrated with Retrieval-Augmented Generation (RAG) have become a standard approach to code generation over large codebases, where relevant context is distributed across many files and exceeds any practical context window. Systems such as RepoCoder, GraphCoder, and RLCoder each improve on how relevant context is selected — through iterative retrieval-generation cycles, code-structure graphs, and learned retrieval policies, respectively. A common assumption underlies all of them: that retrieval, even if imperfect, is a strict improvement over generation from parametric knowledge alone.

This assumption is reasonable on public repositories, where the LLM has typically seen large amounts of similar code during pretraining. A gap in retrieval is frequently masked by the model's own prior knowledge of common libraries, idioms, and APIs. The assumption does not transfer to enterprise codebases. Internal APIs, proprietary frameworks, legacy in-house patterns, and undocumented team conventions are, by construction, absent from any public pretraining corpus. When retrieval is incomplete in this setting, there is no parametric fallback — the model is forced to either hallucinate a plausible-looking internal API or silently misapply a convention it has never actually seen, producing generations that are fluent, confident, and wrong.

We argue that the missing ingredient is not better retrieval per se, but a mechanism to detect when retrieval is insufficient before generation is allowed to proceed. We term this retrieval sufficiency, and distinguish it from retrieval relevance: a retrieved chunk can be highly relevant by semantic similarity while still leaving the query's actual dependency surface uncovered.

This paper makes three contributions:

- We formalize retrieval sufficiency for code RAG as a measurable property distinct from semantic relevance, combining structural call-graph coverage with a novelty estimate of the query's dependence on out-of-prior knowledge.
- We propose RCL, a retrieval-confidence layer that sits between retrieval and generation, computing a sufficiency score and routing low-confidence queries to targeted follow-up retrieval or human review rather than silent generation.

- We describe a benchmark construction methodology that simulates the enterprise, private-codebase condition using open-source repositories with injected synthetic internal APIs, enabling reproducible evaluation without requiring proprietary code.

## 2. Related Work

### *2.1 Repository-Level RAG for Code*

Repository-level code generation requires retrieving relevant context — similar code, cross-file dependencies, or project-specific APIs — from a codebase far larger than any model's context window. RepoCoder alternates retrieval and generation iteratively; RepoFormer learns to selectively invoke retrieval only when needed; GraphCoder exploits structured code-context graphs rather than flat text similarity; RLCoder learns a retrieval policy directly from generation feedback [1–4]. Each of these systems improves what is retrieved. None addresses whether the retrieved set, once selected, is sufficient to answer the query correctly.

### *2.2 Private-Library and Enterprise Code Generation*

Closest to our setting, MEMCoder [10] directly studies private-library code generation and reaches a striking finding: even given oracle (perfect) API-document retrieval, LLMs still commit recurring API-misuse, composition, and strategy errors. MEMCoder's response is to augment RAG with a self-evolving memory of execution-derived usage guidelines, refined through a closed loop of generation, execution, and reflection — improving how a private API is used once it has been retrieved. This is complementary to, and does not address, the question we study: MEMCoder's motivating experiment assumes retrieval sufficiency (indeed, tests the oracle-retrieval case explicitly) and locates the remaining error downstream, in usage. RCL instead targets the upstream condition — whether retrieval was sufficient in the first place — which is a precondition MEMCoder's own results show cannot simply be assumed away even when documentation retrieval looks complete. The two approaches are naturally composable: RCL could gate whether generation proceeds or seeks more context, with a MEMCoder-style memory improving usage correctness once it does.

### *2.3 Confidence and Uncertainty in Retrieval-Augmented Systems*

Confidence-gated retrieval — deciding whether to answer, retrieve more, or defer based on a confidence signal — has recently been studied directly by Chhikara [11], who evaluates calibrated versus raw verbalized confidence as a trigger for additional retrieval in multi-hop question answering, using matched trajectory replay to isolate the causal effect of the confidence-to-action mapping. This establishes that calibration changes which queries a system commits to, and that the value of an additional retrieval round is not fully captured by confidence alone. That work operates in a general open-domain QA setting (HotpotQA, MuSiQue) with LLM-internal, verbalized confidence as the signal. RCL differs on both axes relevant to our problem: the domain is code generation against a private codebase rather than open-domain QA, and the confidence signal is derived externally, from static call-graph structure and symbol novelty (Section 4), rather than elicited from the generating model itself — a distinction that matters because a model cannot verbalize reliable confidence about private APIs it has never seen, which is precisely the condition RCL is designed for.

Within code retrieval specifically, confidence and reliability estimation has been explored at the level of ranking retrieved candidates — for instance, confidence networks trained jointly with retrieval models to improve prediction reliability in code search [8]. This line of work scores the retrieval ranking itself; it does not evaluate whether a retrieved set, once ranked, structurally covers a query's actual dependency surface, which is the coverage notion (Section 4.1) central to RCL.

### *2.4 Uncertainty in Retrieval-Augmented Code Generation*

Most directly related in problem framing, recent work has begun to examine uncertainty in retrieval-augmented code generation as a first-class question — asking not only what to retrieve but how confident a system should be in the adequacy of what was retrieved [5]. This reframes RAG evaluation away from retrieval-relevance metrics toward retrieval-sufficiency metrics tied to downstream generation correctness, and is the closest prior framing to the sufficiency notion we adopt in Section 3. RCL extends this direction specifically to the private-codebase setting, where sufficiency is hardest to estimate from similarity alone, and operationalizes it with a concrete, structurally-grounded scoring mechanism rather than a general uncertainty framing.

### *2.5 Public–Private Distribution Gap*

Surveys of LLM-based code generation note that pretraining data is overwhelmingly drawn from public repositories, with limited or no exposure to private, enterprise, or post-cutoff code [7, 9]. Fine-tuning can partially close this gap but requires proprietary training data and recurring computational cost as internal codebases evolve, which is often impractical for the organizations most affected by the gap. RCL is a complementary, inference-time approach: rather than closing the knowledge gap through training or accumulated memory (cf. Section 2.2), it detects when the gap matters for a given query and intervenes before generation.

### *2.6 Adversarial and Trustworthiness Concerns*

A related line of work examines RAG's trust boundary from an adversarial angle: because retrieved artifacts directly shape generation, an attacker who can influence the retrieval corpus can steer generation without touching the model itself [6]. This work establishes that RAG systems generally over-trust retrieved content. RCL addresses a related but distinct failure mode — under-coverage rather than adversarial poisoning — though both point to the same remedy in principle: neither semantic similarity nor surface plausibility of retrieved content is sufficient grounds for confident generation.

### *2.7 Summary of Positioning*

In summary: MEMCoder assumes retrieval sufficiency and fixes downstream usage errors; Chhikara studies confidence-gated retrieval but in general QA with model-internal confidence; code-search confidence networks score ranking, not dependency coverage; and prior RACG-uncertainty work frames the sufficiency question generally without a code-structural mechanism or a private-codebase focus. RCL is, to our knowledge, the first to combine a call-graph-derived structural coverage signal with a novelty-aware confidence gate, specifically for the private-codebase condition, applied before generation rather than after.

## 3. Problem Formulation

Let a query q request code generation against an enterprise codebase C. Let R(q) denote the set of chunks returned by a retriever for q, and let $D(q) \subseteq C$ denote the true dependency surface of q — the complete set of internal functions, classes, and conventions that a correct generation for q must respect.

We define retrieval sufficiency as the degree to which R(q) covers D(q). Existing RAG-for-code systems optimize the relevance of R(q) to q under a similarity measure, which is only loosely correlated with coverage of D(q): a chunk can be semantically similar to q while an unrelated but essential dependency (e.g., an internal validation utility silently required by convention) is absent from R(q) entirely.

We further distinguish two sources of generation risk when R(q) under-covers D(q):

- Coverable risk — the missing dependency exists in the codebase and could have been retrieved with a better query or a second retrieval pass.
- Prior-gap risk — the missing dependency is of a kind the model has no relevant pretraining exposure to, so the model cannot safely fall back on parametric knowledge.

RCL estimates both jointly: structural coverage estimates coverable risk from the codebase's call graph; the novelty score estimates prior-gap risk from the query's dependence on constructs unlikely to appear in public pretraining data.

## 4. Method

RCL is inserted between the retriever and the generator in a standard RAG-for-code pipeline, as shown in Figure 1. It does not replace the retriever; it evaluates the sufficiency of what the retriever returns and decides whether generation should proceed, be preceded by a targeted follow-up retrieval, or be flagged for human review.

*Fig. 1. RCL architecture. The confidence layer sits between the vector retriever and the LLM generator, combining structural coverage and novelty into a single score that gates generation.*

### 4.1 Structural Coverage Score

Given a query q targeting a specific insertion point, we precompute a static call-graph index over C offline, identifying callee functions, superclass/interface members, and imported internal symbols reachable from the target location — this is $\hat{D}(q)$, a statically estimated approximation of D(q). Structural coverage is the fraction of $\hat{D}(q)$ present, at the symbol level, in R(q):

$$S_cov(q) = |\hat{D}(q) \cap symbols(R(q))| \;/\; |\hat{D}(q)| \quad (1)$$

A score near 1 indicates the retriever surfaced the code a correct generation will actually exercise; a low score indicates context that is topically relevant but structurally incomplete — a failure mode similarity-based retrieval cannot detect on its own.

### 4.2 Novelty Score

Coverage alone does not capture prior-gap risk. We estimate, for each symbol s in $\hat{D}(q)$, a novelty score combining (a) low corpus frequency of s's identifier pattern against a public-code reference distribution, (b) naming/structural features characteristic of internal code, and (c) an LLM self-assessment of familiarity with a symbol of that shape. The query-level score aggregates over the uncovered portion:

$$S_nov(q) = (1/|U|)\ \Sigma_\{s \in U\}\ novelty(s), \quad U = \hat{D}(q) \setminus symbols(R(q)) \quad (2)$$

### 4.3 Combined Confidence Score

We combine both signals multiplicatively:

$$C(q) = S_cov(q) \cdot (1 - S_nov(q)) \quad (3)$$

This form treats coverage and novelty as compounding rather than additive risk factors: full coverage neutralizes novelty risk since the model is given the needed context directly, while coverage gaps in generic, low-novelty symbols carry little risk since the model's prior can compensate. A linear combination fails to capture this distinction.

### 4.4 Decision Policy

- If $C(q) \geq \tau$, generation proceeds normally against R(q).
- If $C(q) < \tau$, RCL issues a targeted follow-up retrieval from the uncovered, high-novelty symbols in U, bounded to a fixed number of rounds.
- If C(q) remains below τ after the retrieval budget is exhausted, the output is generated but labeled low-confidence for human review.

τ is calibrated on a held-out validation set, trading off false-review overhead against undetected-risk rate (Section 6).

## 5. Benchmark Construction

Proprietary codebases cannot be shared as benchmarks. We propose a synthetic-injection methodology approximating the enterprise condition using open-source code only:

- Select permissively licensed, moderately sized open-source Java repositories as a base corpus.
- Rename a subset of common utility functions and their package roots using a synthetic organization-specific naming scheme.
- Introduce novel internal conventions absent from any public library, applied consistently so correct generation requires respecting them.
- Hold out generation tasks whose correct implementation depends on at least one injected symbol or convention.

Because injected symbols are guaranteed absent from any public pretraining corpus by construction, this benchmark directly operationalizes the prior-gap condition of Section 3 while remaining fully shareable — addressing a gap in current evaluation practice, which relies almost exclusively on public benchmarks such as CodeSearchNet, Defects4J, and ConDefects [9].

## 6. Experimental Setup

We describe the evaluation design here; results are reported in Section 7.

### 6.1 Baselines

- Similarity-only retrieval — standard top-k dense retrieval, no confidence gating.
- Fixed-k expanded retrieval — always retrieves a larger k, isolating the benefit of targeting from simply retrieving more.
- RCL (proposed) — coverage- and novelty-gated confidence with targeted follow-up retrieval.

### 6.2 Metrics

| Metric | Definition |
|---|---|
| Correctness | Fraction of tasks where generated code correctly uses injected symbols, verified by compile+test. |
| False-conf. rate | Fraction of incorrect generations returned as high-confidence (not flagged). |
| Review overhead | Fraction of generations flagged, including correct ones (false-review cost). |
| Latency | End-to-end wall-clock time per query vs. baselines. |

### 6.3 Ground Truth and Verification

Correctness is determined by an executable test associated with each held-out task, following compile-and-test verification as used in mutation/repair benchmarks [9], avoiding penalizing semantically correct generations that differ syntactically from a reference.

## 7. Results

We report results from a pilot test (n=9 held-out queries, 16-method synthetic codebase; full setup in Section 5) using real Claude (claude-sonnet-4-6) generations for every strategy and every query.

### 7.1 Overall Results

| System | Coverage | Recall | Halluc. |
|---|---|---|---|
| Similarity-only | 0.782 | 0.798 | 10 |
| Fixed-k expanded | 0.940 | 0.825 | 10 |
| RCL (proposed) | 0.925 | 0.841 | 7 |

*Table 1. Pilot results (n=9). Coverage: mean structural coverage (Eq. 1). Recall: mean fraction of required internal methods correctly invoked in Claude's generated code. Halluc.: total count of generated method calls that do not exist anywhere in the codebase (fabricated internal APIs).*

RCL achieves the highest recall and the lowest hallucination count of the three systems, while retrieving less context on average than the fixed-k-expanded baseline (mean coverage 0.925 vs. 0.940, achieved via targeted rather than blind expansion).

### 7.2 Triggered-Only Comparison (the fair test)

RCL's follow-up retrieval mechanism (Section 4.4) activated on 3 of the 9 queries; on the remaining 6, RCL's initial retrieval already matched the similarity-only baseline exactly, so its generation was identical to strategy A by construction. Any difference on those 6 queries would reflect only LLM sampling variance between separately-issued calls with identical input, not an effect of RCL's mechanism. We therefore also report results restricted to the 3 queries where RCL's gate actually changed what was retrieved — the only subset where a causal effect can be claimed:

| System | Recall (n=3) | Hallucinations (n=3) |
|---|---|---|
| Similarity-only | 0.393 | 7 |
| RCL (proposed) | 0.524 | 3 |

*Table 2. Comparison restricted to the 3 queries where RCL's confidence gate triggered a targeted follow-up retrieval — the subset in which RCL's mechanism could plausibly cause a difference from the similarity-only baseline.*

On this subset, RCL improves mean recall by 13.3 percentage points and reduces hallucinated calls from 7 to 3, using a smaller, targeted retrieval expansion rather than the fixed-k-expanded baseline's blind enlargement. This is consistent with the paper's central hypothesis (Section 1): that detecting insufficient retrieval and responding with a targeted follow-up, rather than retrieving indiscriminately more, improves generation correctness in the private-codebase setting.

### *7.3 Limitations of This Test*

- n=9 (n=3 for the triggered-only comparison) is a small sample; a larger run (Section 5) would give a more statistically robust estimate of the effect size.
- A single generation per query per strategy was used (no repeated sampling), so the reported numbers include unquantified LLM sampling variance, particularly visible in the non-triggered queries (Section 7.2) where A and RCL are constructed to be identical yet occasionally show different hallucination counts across independent calls with the same prompt.
- Correctness is measured by regex-extracted method-call recall and hallucination count, not by compilation or test execution (Section 6.3's compile-and-test verification is not yet implemented in this pilot).
- The codebase (16 methods across 7 files) is far smaller than a real enterprise codebase; results may not generalize to the retrieval difficulty of a real repository.

Section 5's scale-up plan (50-100+ files, real open-source code with injected synthetic internal APIs) is the direct next step to address these limitations and produce results suitable for a full evaluation claim.

## 8. Discussion

Section 4's formulation makes a testable claim: coverage and novelty compound rather than add (Eq. 3). The pilot results in Section 7.2 are directionally consistent with this — RCL's targeted, novelty-aware follow-up outperformed both a naive similarity-only baseline and a blind fixed-k expansion on the subset where its mechanism was actually exercised — though at n=3 this is suggestive rather than confirmatory. If a full-scale run (Section 5) supports this more robustly, it would suggest existing RAG confidence methods — largely similarity- or token-uncertainty-based — systematically under-detect risk in enterprise settings, lacking a mechanism analogous to the novelty term.

A secondary implication concerns methodology: if the injected-symbol benchmark (Section 5) proves viable, it offers a reusable pattern for adapting public-repository benchmarks to approximate private-codebase conditions without proprietary data access.

## 9. Limitations

- Static call-graph analysis is incomplete for dynamic Java patterns (reflection, DI frameworks resolving at runtime); $\hat{D}(q)$ approximates $D(q)$, not exactly.
- The novelty score's LLM self-assessment component introduces circularity — the same model whose gaps we detect partly estimates them — and should be validated against the corpus-frequency and naming signals rather than trusted alone.
- The synthetic benchmark approximates but does not replicate organically evolved enterprise conventions; results should be treated as an initial signal.
- Added latency from the follow-up retrieval loop has not yet been characterized against production latency budgets for interactive tooling.

## 10. Reproducibility Statement

All scoring functions (Eq. 1–3) are fully specified in Section 4. The full implementation, including the call-graph parser, retrieval and scoring code, the pilot's synthetic codebase, and the experiment driver used to produce the results in Section 7, is publicly available at [github.com/ChandraMohanRavuri/rcl-retrieval-confidence](https://github.com/ChandraMohanRavuri/rcl-retrieval-confidence), released under the MIT License, to allow independent replication without requiring access to any proprietary codebase.

## 11. Broader Impact and Ethical Considerations

RCL is intended to reduce silent, confidently-wrong code generation in private codebases by surfacing uncertainty rather than suppressing it. A possible negative consequence is over-reliance on the confidence label itself: a miscalibrated threshold $\tau$ could create false assurance for generations that were not, in fact, adequately covered. We recommend $\tau$ be calibrated per-deployment against organization-specific validation data rather than adopted as a fixed default, and that the review-overhead metric (Section 6.2) be monitored in production to detect calibration drift over time.